\documentclass[12pt]{article}
\pdfoutput=1
\usepackage{epsf,amsfonts,amssymb,epsfig,amsmath,mathtools,graphics,slashed}
\usepackage{hyperref,hep,graphicx,subfig}
\usepackage{rotating}

\newcommand{\nn}{\notag \\}

\makeatletter

\@addtoreset{equation}{section}
\makeatother

\begin{document}

\begin{titlepage}

\vfill

\begin{flushright}
Imperial/TP/2016/JG/03\\
DCPT-16/33
\end{flushright}

\vfill

\begin{center}
   \baselineskip=16pt
   {\Large\bf Thermal backflow in CFTs}
  \vskip 1.5cm
  \vskip 1.5cm
Elliot Banks$^1$, Aristomenis Donos$^2$, Jerome P. Gauntlett$^{1,3}$\\ Tom Griffin$^1$ and Luis Melgar$^1$\\
     \vskip .6cm
      \begin{small}
      \textit{$^1$Blackett Laboratory, 
  Imperial College, London, SW7 2AZ, U.K.}
        \end{small}\\
         \vskip .6cm
      \begin{small}
      \textit{$^2$Centre for Particle Theory and Department of Mathematical Sciences\\Durham University, Durham, DH1 3LE, U.K.}
        \end{small}\\
       \vskip .6cm
      \begin{small}
      \textit{$^3$School of Physics, Korea Institute for Advanced Study, Seoul 130-722, Korea}
        \end{small}\\

\end{center}

\vfill

\begin{center}
\textbf{Abstract}
\end{center}
\begin{quote}
We study the thermal transport properties of general conformal field theories (CFTs)
on curved spacetimes in the leading order viscous hydrodynamic limit. 
At the level of linear response, we show
that  the thermal transport is governed by a system of forced linearised Navier-Stokes equations on a curved space. Our setup includes CFTs in flat spacetime that have been deformed by spatially dependent and periodic local temperature variations or strains that have been applied to the CFT, and hence is relevant to CFTs arising in condensed matter systems at zero charge density. We provide specific examples of deformations which lead to thermal backflow driven by a DC source: that is, the thermal currents locally flow in the opposite direction to the applied DC thermal source. We also consider thermal transport for relativistic quantum field theories that are not conformally invariant.
\end{quote}

\vfill

\end{titlepage}

\setcounter{equation}{0}
\section{Introduction}

A wide variety of strongly correlated states of matter are expected to display collective behaviour 
described by viscous hydrodynamics. This occurs on time scales when the momentum preserving self interactions 
of the strongly coupled matter dominate over momentum dissipating processes such as the scattering 
with phonons. For some further discussion, including some experimental realisations
in graphene and other materials, we refer to 
\cite{0038-5670-11-2-R07,
:/content/aip/journal/jap/69/2/10.1063/1.347315,
PhysRevB.51.13389,
1997PhRvB..56.8714D,
Herzog:2007ij,
Forcella:2014gca,
2016NatPh..12..672L,
2016Sci...351.1055B,
2016Sci...351.1058C,
2016Sci...351.1061M,
Lucas:2015sya,
2016arXiv160700986F,
2016arXiv160707269G}.  
It is has recently been emphasised that for matter at finite charge density a directly verifiable macroscopic signature of viscous flows is provided by the phenomenon of electric current backflow \cite{2016NatPh..12..672L,2016Sci...351.1055B}. 
That is, for suitable set-ups the application of an external electric field leads to a fluid flow that produces an
electric current which flows, locally, in the opposite direction to the applied field. 

Here we want to discuss thermal backflow. 
In this case a local heat current flows in the opposite direction to that of an applied external
temperature gradient 
and in principle can occur in the absence of charge carriers. 
While electric backflow can be caused both by viscous effects and by spatially modulated regions of charge density (``charge puddles"), thermal backflow would be caused purely by viscous effects of the fluid.
For matter at finite charge density, both are special cases of the more general phenomenon of thermoelectric current backflow.  
 
In this paper we initiate a study of thermoelectric current backflow for relativistic quantum field theories,
focussing on conformal field theories (CFTs).
More specifically, we will investigate the possibility of thermal backflow by applying an external
DC thermal gradient to CFTs
at finite temperature and vanishing charge density. We then calculate the local currents that are produced at the level of linear response by solving leading order viscous hydrodynamic equations.  We are interested in studying this phenomenon for infinite systems. Thus, in order to
get a finite DC response we will need a set-up in which the total momentum is a not a conserved quantity, or 
phrased differently, momentum dissipates in the bulk of the CFT.
This should be contrasted with other setups where a finite DC response arises because 
one imposes no-slip or other momentum dissipating boundary conditions on the electronic fluid
in a finite volume, as in some of the discussion in \cite{2016NatPh..12..672L,2016Sci...351.1055B}, for example.
A natural way to achieve this is to consider CFTs in Minkowski spacetime that are then deformed
by marginal or relevant operators that explicitly break the translation invariance of the CFT. Interestingly, this is precisely 
the set-up that has received much attention in the AdS/CFT correspondence via the construction of
black holes called  ``holographic lattices"  \cite{Hartnoll:2012rj,Horowitz:2012ky,Donos:2012js,Chesler:2013qla,Donos:2013eha,Andrade:2013gsa}.
 
Here we will focus on the universal class of deformations that arise from placing the CFT on a curved geometry with spacetime metric $g_{\mu\nu}(x)$. We assume that the metric is time independent, {\it i.e.} it has a timelike Killing vector $\partial_t$, corresponding to a CFT in local thermal equilibrium. The metric $g_{\mu\nu}(x)$ can also be viewed 
as parametrising spatially dependent sources for the stress tensor of the CFT.
These deformations include applying strains, thermal gradients as well as
sources for local rotations to a CFT in flat spacetime, for example. 
In thinking of potential applications to real materials we can envisage applying such
deformations to a plasma that has arisen from some underlying collective behaviour. 
For example, we note that there has been extensive work on studying the behaviour of strained graphene, {\it e.g.} \cite{RevModPhys.81.109,PhysRevB.81.035411,C5NR07755A} and it is also worth highlighting the exceptional
thermal conductivity properties of graphene \cite{thcond}.

We will study the linear response of the deformed CFTs at vanishing charge density after applying 
an external thermal gradient source, possibly time dependent, in the hydrodynamic limit, 
$\epsilon=k/T <<1$, where $k$ is the largest wavenumber associated with the deformations.
For the special case of CFTs with holographic duals it has been shown that there is a universal connection between
thermal DC conductivity and Navier-Stokes equations on black hole horizons \cite{Donos:2015gia}. Using these results, it was recently shown for holographic lattices in the hydrodynamic limit  that the local heat current that is produced by a thermal source can be obtained by solving a system of linearised, forced Navier-Stokes equations on a curved manifold fixed by the metric $g_{\mu\nu}$ \cite{Banks:2016krz}.
In this paper we will show that this result is much more general, applying also to general CFTs without holographic duals. We will also show how it also arises for non-conformally invariant relativistic field theories.

To illustrate thermal backflow for a DC source we will study static metrics with spatial sections that are conformally flat, with the conformal factor a periodic function of the spatial coordinates. This corresponds to applying an isotropic periodic strain
to the CFT.  After applying a Weyl transformation it also corresponds to deforming by a spatially modulated energy distribution, or equivalently a spatially modulated local temperature variation. 
For suitably chosen conformal factors, by solving the time-independent Navier-Stokes equations numerically, we are
able to find explicit examples that do indeed exhibit thermal backflow for this setup. 
We emphasise that this thermal backflow arises at the level of the linear 
response to the application of an external DC thermal gradient, and is thus associated with specific two point functions of the stress tensor in the strained CFT.
Moreover, the backflow is due to the spatial inhomogeneities
of the metric on which the CFT lives and this should be contrasted with fluid backflow 
in ordinary fluids at the level of linear response, that is caused by momentum dissipating processes
at boundaries.

We will focus on CFTs in the bulk of the text because their hydrodynamic description depends on fewer parameters.
However, much of our analysis can be straightforwardly 
generalised to arbitrary relativistic quantum field theories and we present some
details in appendix \ref{gencase}. It is interesting that for static metric backgrounds with the time-like Killing vector having constant norm, we also find that the response to a thermal source, possibly time dependent, is again
governed by linearised Navier-Stokes equations. For non-constant norm, we obtain more general equations.

\section{Thermal transport for CFTs in the hydrodynamic limit}\label{sec1}
We consider general CFTs on curved manifolds in $d\ge 2$ spacetime dimensions with metric $g_{\mu\nu}$.
Using the general results of \cite{Baier:2007ix} (see also \cite{Loganayagam:2008is}), we will derive the
leading order viscous hydrodynamic equations relevant for studying thermal transport 
after applying an external thermal gradient source, possibly time dependent, at the level
of linear response. 

For a general CFT we must impose the Ward identities
\begin{align}\label{wids}
D_\mu T^{\mu\nu}=0\,,\qquad T^\mu{}_\mu=0\,.
\end{align}
When $d$ is even we have set the conformal anomaly to zero as it will be higher order in the derivative expansion than
we wish to consider.
In order to obtain a closed set of hydrodynamical equations we need constitutive relations for
the stress tensor. We let $T$ denote the local temperature and introduce the fluid velocity $u^\mu$, satisfying
$u^\mu u_\mu=-1$. Both $T$ and $u^\mu$ can depend on all of the spacetime coordinates, $x^\mu$.
Including the leading order viscous terms we have
\begin{align}
\label{eq:EMTe}
T_{\mu \nu} = P (g_{\mu \nu} + d\, u_{\mu}u_{\nu}) - 2 \eta \sigma_{\mu \nu}\,,
\end{align}
where the shear tensor is given by
\begin{align}\label{shear}
\sigma_{\mu \nu} = D_{(\mu} u_{\nu)} + u_{(\mu} u^{\rho}D_{\rho} u_{\nu)} - (g_{\mu \nu} +u_{\mu}u_{\nu}) \frac{D_{\rho} u^{\rho}}{d-1}\,.
\end{align} 
Conformal invariance fixes the equation of state to be $P=c_{0}\,T^{d}$ and the viscosity to be $\eta=c_{1}\,T^{d-1}$, where $c_{0}$ and $c_{1}$ are dimensionless numbers fixed by the CFT\footnote{In holography we have 
$c_{0}=\frac{4\pi}{d}c_{1}$.}.

Notice that the equations are covariant under Weyl transformations, in which the metric and fluid velocity transform as $g_{\mu\nu}\to e^{2\omega}g_{\mu\nu}$, $u_{\mu}\to e^{\omega} u_{\mu}$, where $\omega$ is an arbitrary function of spacetime coordinates, while the scalars $T$, $P$, $\eta$ 
transform as $T\to e^{-\omega}T$, $P\to e^{-d\,\omega}\,P$ and $\eta\to e^{(-d+1) \omega}\eta$.
We also notice that $u^{\mu}T_{\mu\nu}=-(d-1)Pu_\nu=-\varepsilon \, u_{\nu}$, where $\varepsilon$ is the energy density and we also have 
$\varepsilon+P=sT$.

Introducing a time coordinate via $x^\mu =(t,x^i)$, then the heat current density, or equivalently, momentum current density, of the CFT is given by
the components
\begin{align}\label{qexp}
Q^i=-\sqrt{-g}\,T^i{}_t\,.
\end{align}
Notice that $Q^i$ is invariant under Weyl transformations.
Also, in stationary spacetimes, for which $\partial_t$ is a Killing vector, we deduce that this current is 
conserved $\partial_i Q^i=0$.

To simplify the presentation, we now consider the background metric to be static with line element given by $ds^2=-g_{tt}dt^2+g_{ij}dx^i dx^j$, and $\partial_tg_{tt}=\partial_tg_{ij}=0$. 
This corresponds to studying the CFT in thermal equilibrium, with $g_{tt}$ and $g_{ij}$ parametrising
sources for the stress tensor components $T^{tt}$ and $T^{ij}$, respectively.
It will be convenient to set $g_{tt}=1$ and consider the background metric
\begin{align}\label{met}
ds^2=-dt^2+g_{ij}(x^k)dx^i dx^j\,,
\end{align}
since a non-vanishing $g_{tt}$ can be reinstated by simply performing a Weyl transformation. 
We next consider the spatial metric $ds^2=g_{ij}(x^k)dx^i dx^j$ as a harmonic expansion about
some fiducial metric. If $k$ is the largest wave-number in this expansion, then the hydrodynamic limit
has $\epsilon=k/T<<1$. A concrete example, and one we will focus on, is to take the fiducial metric to be flat space and 
consider $g_{ij}$ to be periodic in the spatial directions. In this case, focussing on a fundamental domain, 
$g_{ij}$ also defines a curved metric on a torus.

We now consider perturbing the CFT by an external
thermal gradient source parametrised by a closed one form $\zeta=\zeta_\mu dx^\mu$.
To study the linear response of the CFT to this source, similar to \cite{Banks:2016krz}, we
consider the following linearised perturbation about the equilibrium configuration. For the metric we take\footnote{Employing the 
coordinate transformation $t= (1+\phi)\bar t$ implies that the linearised perturbed metric is given by 
$ds^{2}=-d\bar t^{2}+g_{ij}(x)dx^{i}dx^{j}-2\bar t\zeta_\mu dx^\mu d \bar t$, which has been used in related contexts \cite{Donos:2015gia}.}.
\begin{align}\label{dcpert}
ds^{2}=-(1-2\phi)\,dt^{2}+g_{ij}(x)dx^{i}dx^{j}\,,
\end{align}
where $\zeta_\mu=\partial_\mu\phi$. 
We now highlight an important aspect of the choice of $\zeta$ and $\phi$. 
To illustrate, we focus on the planar case with $g_{ij}(x)$ periodic in the spatial directions. In this case we can write 
$\zeta=\bar\zeta_i(t) dx^i +dz(t, x)$, or $\phi= \bar\zeta_i(t) x^i +z(t,x)$, where $z(t,x)$ are periodic functions of the $x^i$.
The $\bar\zeta_i$ parametrise the thermal source of most interest. For example, for the DC case, the choice 
$\phi(x)=z(x)$ would just correspond to considering the CFT on a deformed metric still in thermal equilibrium 
(we return to this at the end of the section). On the other hand $\phi= \bar\zeta_i x^i$, with constant $\bar\zeta_i$
corresponds to a constant external thermal gradient source, of strength $\bar\zeta_i$, in the $x^i$ direction\footnote{ 
Note that $\phi(x)=z(x)$ is globally defined and bounded both on the plane and on the torus (i.e. 
associated with a fundamental domain of the background). On the other hand $\phi= \bar\zeta_i x^i$
is globally defined on the plane, but not bounded, and is not a well defined function on the torus. Furthermore,
the one-forms $dz(x)$ and $\bar\zeta_i dx^i$ are cohomologically trivial and non-trivial on the torus, respectively.}.

We consider the perturbed fluid velocity to be
\begin{align}
u_t=-(1-\phi),\qquad u_j=\delta u_j\,.
\end{align}
We vary the local temperature via $T=T_0+\delta T$, where
 $T_0$ is the equilibrium temperature of the CFT. Note that $\phi$, $\delta u_i$ and $\delta T$ all depend
 on $(t,x^i)$; in the planar case they are taken to be periodic functions of the $x^i$.
 If $\omega$ is a characteristic frequency then we should demand that $\omega/T_0<<1$ in
 addition to $k/T_0<<1$, in order to stay in the hydrodynamic limit.  

After substituting into \eqref{eq:EMTe} we find that the stress tensor takes the form 
\begin{align}\label{stressex}
T_{tt}&=c_{0}\,(d-1)\,T^{d}_{0}(1-2\phi)+c_{0}d\,(d-1)\,T_{0}^{d-1}\,\delta T,\nn
T_{ti}&=-c_{0}d\,T^{d}_{0}\,\delta u_i\,,\nn
T_{ij}&=c_{0}T^{d}_{0}\,g_{ij}+c_{0}\,d\,T^{d-1}_{0}\,\delta T\, g_{ij}-2 c_{1}T^{d-1}_{0}\,\left(\nabla_{\left(i\right.} \delta u_{\left. j\right)}-\frac{g_{ij}}{d-1} \nabla_{k}\delta u^{k}\right)\,,
\end{align}
where here, and below, the covariant derivative $\nabla$ is now with respect to $g_{ij}$.
The Ward identities \eqref{wids} then give the following linearised, forced
Navier-Stokes equations for $\delta u_i$ and $\delta T$:
\begin{align}\label{stokesgeneral}
T_0\,\partial_{t}\delta u_{i}-2\frac{c_{1}}{d\,c_{0}}\,\left( \nabla^{j}\nabla_{\left(j\right.} \delta u_{\left. i\right)}-\frac{1}{d-1}\,\nabla_{i}\,\nabla_{j}\delta u^{j} \right)+\nabla_{i}\delta T&=T_{0} \zeta_{i}\,,\notag\\
(d-1)T_0^{-1}\partial_t\delta T+\nabla_{i}\delta u^{i}&=0\,.
\end{align}
Furthermore, the heat current \eqref{qexp} now reads
\begin{align}\label{qexp2}
Q^i=c_{0}dT^{d}_{0}\sqrt{g}\,\delta u^{i}=T_0s_0\sqrt{g}\,\delta u^{i}\,.
\end{align}

The system of equations \eqref{stokesgeneral} is the key result of this section. 
Observe that they only depend on the one-parameter of the CFT, $c_1/(d c_0)$, which is just $\eta_0/s_0$. We also note
that $\zeta_t$ does not enter these equations.
When we set all time derivatives to zero, which is appropriate for studying thermal DC response, we have an 
incompressible fluid $\nabla_{i}\delta u^{i}=0$. We will refer to the time-independent equations as Stokes equations.

We conclude this section with a few general comments. 
We first make some observations about conserved currents for general relativistic 
field theories satisfying
the Ward identity $D_\mu T^{\mu\nu}=0$ on curved manifolds, setting
to zero the thermal sources ({i.e.} $\phi=0$).
Contracting with an arbitrary vector $k^\mu$ we obtain 
\begin{align}\label{genres}
D_\mu\left(T^\mu{}_\nu k^\nu\right)
=\tfrac{1}{2}{\cal L}_k g_{\mu\nu}T^{\mu\nu}\,,
\end{align}
where ${\cal L}$ is the Lie derivative. We immediately see that if $k$ is a 
Killing vector then $T^\mu{}_\nu k^\nu$ is a conserved current. For a CFT this is also true if $k$
is a conformal Killing vector, satisfying ${\cal L}_k g_{\mu\nu}\propto g_{\mu\nu}$.
Thus, in order to have momentum dissipation in the spatial directions, we should only consider background metrics without (conformal) Killing vectors, apart from $\partial_t$. Equivalently, for a CFT, the metric should not be related by a Weyl transformation to a metric with additional Killing vectors. If we let $k=\partial_i$ and assume that it is not a (conformal) Killing vector, then there is no conserved momentum in the $x^i$ direction.  In this case, if we 
consider perturbing around thermal equilibrium, \eqref{genres} might be viewed as saying that
momentum is being dissipated by the non-vanishing of
$\partial_i g_{\mu\nu} \delta T^{\mu\nu}$. This can be contrasted with the work of \cite{Hartnoll:2007ih} who, instead, modify the Ward identities in order to achieve momentum dissipation.

We now consider a stationary metric $g_{\mu\nu}$ and assume that
$k^\mu$ is a Killing vector (or conformal Killing vector if we have a CFT), in addition to $\partial_t$. After considering a DC perturbation \eqref{dcpert}, with all time derivatives vanishing, from the Ward identity we deduce that
\begin{align}
\frac{1}{\sqrt{g}}\partial_i\left(\sqrt{g}T^i{}_\mu k^\mu\right)=-(k^i\zeta_i)T^t{}_t\,.
\end{align}
After integrating over the spatial directions, the left hand side vanishes\footnote{With appropriate boundary conditions imposed for non-compact spaces.} and hence so does the right hand side. Thus, we have deduced, just from the Ward identity (i.e. independent of the constitutive relations), that
if there are any (conformal) Killing vectors over and above $\partial_t$, then the DC response is not well defined in the direction
$k^i\zeta_i$. More physically, there will be a delta function at zero frequency in the AC response.

In studying DC response for background metrics as in \eqref{met}, we are thus only interested in 
spatial metrics $g_{ij}dx^i dx^j$ without Killing vectors. The solutions to the Stokes equations ({\it i.e.} \eqref{stokesgeneral} with $\partial_t=0$) are then unique \cite{Donos:2015gia,Banks:2015wha}
up to an undetermined constant, the zero-mode of $\delta T$. Physically, this zero mode can be fixed by demanding that
when $\zeta_i=\delta u_i=0$ the full stress tensor of the CFT is not modified. In any event, this zero mode does not
affect the local heat current response given in \eqref{qexp2}.

The final comment relates to the closed one form 
source $\zeta$ in the DC context. 
For the periodic, planar case we again write
$\zeta=\bar\zeta_i dx^i +dz(x)$, where $\bar\zeta_i dx^i$, with constant $\bar \zeta_i$, parametrise the DC thermal source of most interest,
and $z(x)$ is an arbitrary periodic function which can be dealt with exactly. 
Indeed as noted in \cite{Donos:2015gia,Banks:2015wha} if $\zeta=dz(x)$, associated with $\phi=z(x)$, we can solve the Stokes equations
with $\delta u_i=0$ and $\delta T=T_0 z$, giving rise to a simple response to the full stress tensor with no heat flow. 
Note that we cannot take the solution $\delta u_i=0$ and $\delta T=T_0 \phi$ when $\phi(x)=\bar\zeta_i x^i$ since we have demanded
that $\delta u_i$ and $\delta T$ are periodic functions\footnote{Note that an alternative approach, in this DC context, would have been to allow
non-periodic perturbations $\delta T$ instead of non-periodic functions $\phi(x)$.}.

\section{Thermal backflow}
We now consider specific background static metric deformations of the form \eqref{met}, parametrised by $g_{ij}(x^k)$,
that lead to thermal backflows driven by external DC thermal gradients, in the hydrodynamic limit. We will assume that we have a planar spatial topology
with $g_{ij}$ a periodic function of the spatial coordinates. 
For a given $g_{ij}$ we want to numerically solve the Stokes equations ({\it i.e.} \eqref{stokesgeneral} with $\partial_t=0$), effectively on a torus,
and then obtain the local heat current density, $Q^i(x)$, at leading order in $k/T$,
using \eqref{qexp2}.

For simplicity we will assume that the deformation is periodic in each of the spatial directions with the same period, $L\equiv 2\pi/k$, with $\epsilon=k/T<<1$. For the numerics we eliminate the dimensionful quantity $L$ by defining new coordinates via
$x^{i}=L\hat x^{i}$ with the $\hat x^i$ having unit period. It is convenient to introduce dimensionless variables via\footnote{Note that $p$, here should not be confused with the pressure, $P$, of the background CFT appearing in \eqref{eq:EMTe}.}
\begin{align}
v_{i}=\delta u_{i},\qquad  p=\frac{d\,c_{0}}{c_{1}}\,L\delta T,\qquad \hat\zeta_i=\frac{d\,c_{0}}{c_{1}}\,L^2T_{0}\,\zeta_{i}\,.
\end{align}
Then in the hatted coordinates the linear Stokes equations coming from \eqref{stokesgeneral} take the dimensionless form
\begin{align}\label{stokesgeneral2}
-{2}\,{\nabla}^{i}{ \nabla}_{\left(i\right.} v_{\left.j\right)}=\hat \zeta_{j}-\partial_{j} p\,,\qquad \nabla_i  v^i=0\,,
\end{align}
where here we are raising indices with respect to the metric $g_{ij}$ and $\nabla$ is the associated covariant derivative.
In the new variables it is
natural to define the heat current density
\begin{align}
\hat Q^i&\equiv \sqrt{ g} g^{ij} v_j
=\frac{1}{c_{0}dT_{0}^{d}}Q^i\,.
\end{align}
Writing $Q^i=T_0\kappa^{ij}\zeta_j$, where $\kappa$ is the thermal conductivity matrix, then we have
$\hat Q^i=(c_1/c_0 d) (T_0\kappa^{ij}/s_0)\epsilon^2\zeta_j$, where $s_0$ is the entropy density. This displays the
fact that $(T_0\kappa^{ij}/s_0)$ is of order $\epsilon^{-2}$ as pointed out in \cite{Banks:2016krz}.

\begin{figure}[h!]
\centering
\includegraphics[width=0.5\textwidth]{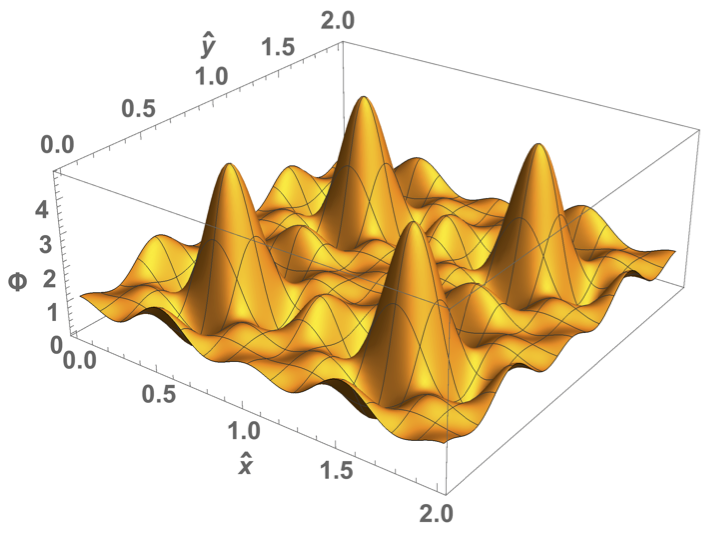}
\caption{A plot of the function $\Phi$ which determines the static metric deformation of the CFT with $g_{tt}=1$ and $g_{ij}=\Phi\delta_{ij}$. Note that we have plotted twice the period in both spatial directions. This specific choice of $\Phi$ is as in \eqref{fdef} with $\alpha=0.98$ and $\beta=0.3$ and gives rise to thermal backflow as shown in figure \ref{defphiresults}.
\label{defphi}}
\end{figure}

To illustrate examples of backflow, we now restrict to CFTs 
with metric deformations given by\footnote{Note that for this choice of metric, \eqref{genres} with $k=\partial_i$
gives $\nabla_\mu \delta T^{\mu}{}_i=-\tfrac{1}{2}(\partial_i\ln \Phi)\delta T^{t}{}_{t}$, revealing the origin of
momentum non-conservation in this setting.} 
\begin{align}
g_{ij}=\Phi\delta_{ij},\qquad \Phi >0\,.
\end{align}
By solving the Stokes equations \eqref{stokesgeneral2} numerically, 
we find that various choices of $\Phi$ lead to thermal backflow. To be specific
we discuss the special case of CFTs in two spatial dimensions and set $d=3$. We 
present some results for the specific choice 
\begin{align}\label{fdef}
\Phi=\alpha+\frac{\beta}{N}\,\sum_{a,b=-N}^{N}e^{i\,2\pi\,(a\,(\hat x-1/2)+b\,(\hat y-1/2))}\,.
\end {align}
Moreover, we restrict to the specific case of $N=2$ and consider varying $\alpha$ and $\beta$.
We have plotted $\Phi$ for the specific case of $\alpha=0.98$ and $\beta=0.3$ in figure \ref{defphi}.
We apply a constant DC thermal gradient just in the $\hat x$ direction with $\hat\zeta =d\hat x$.
For various choices of $\alpha,\beta$ we then numerically solve the Stokes equations \eqref{stokesgeneral2}, as described in appendix \ref{appnum}, to extract $\hat Q^i(\hat x)$ and $p(\hat x)$.

\begin{figure}[h!]
\centering
\includegraphics[width=0.45\textwidth]{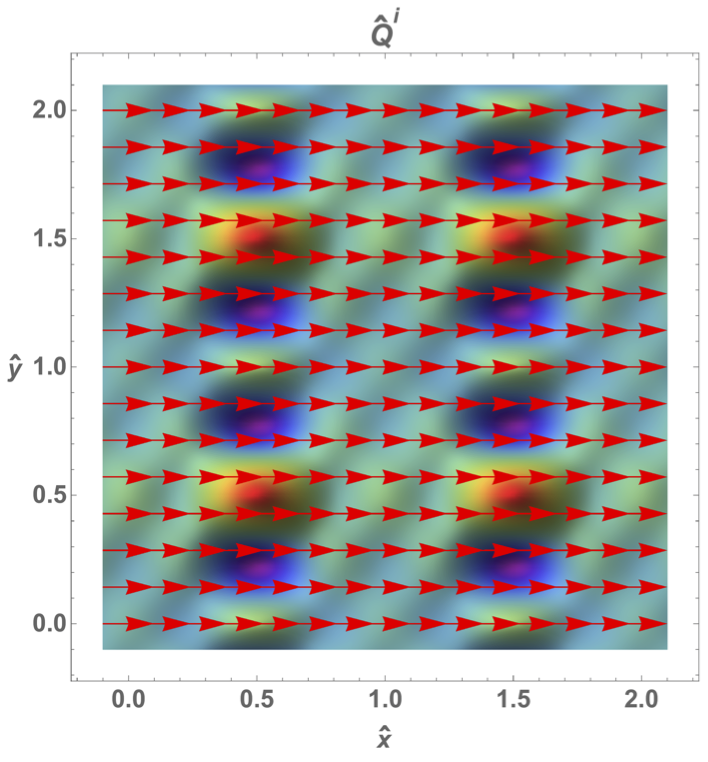}\qquad
\includegraphics[width=0.45\textwidth]{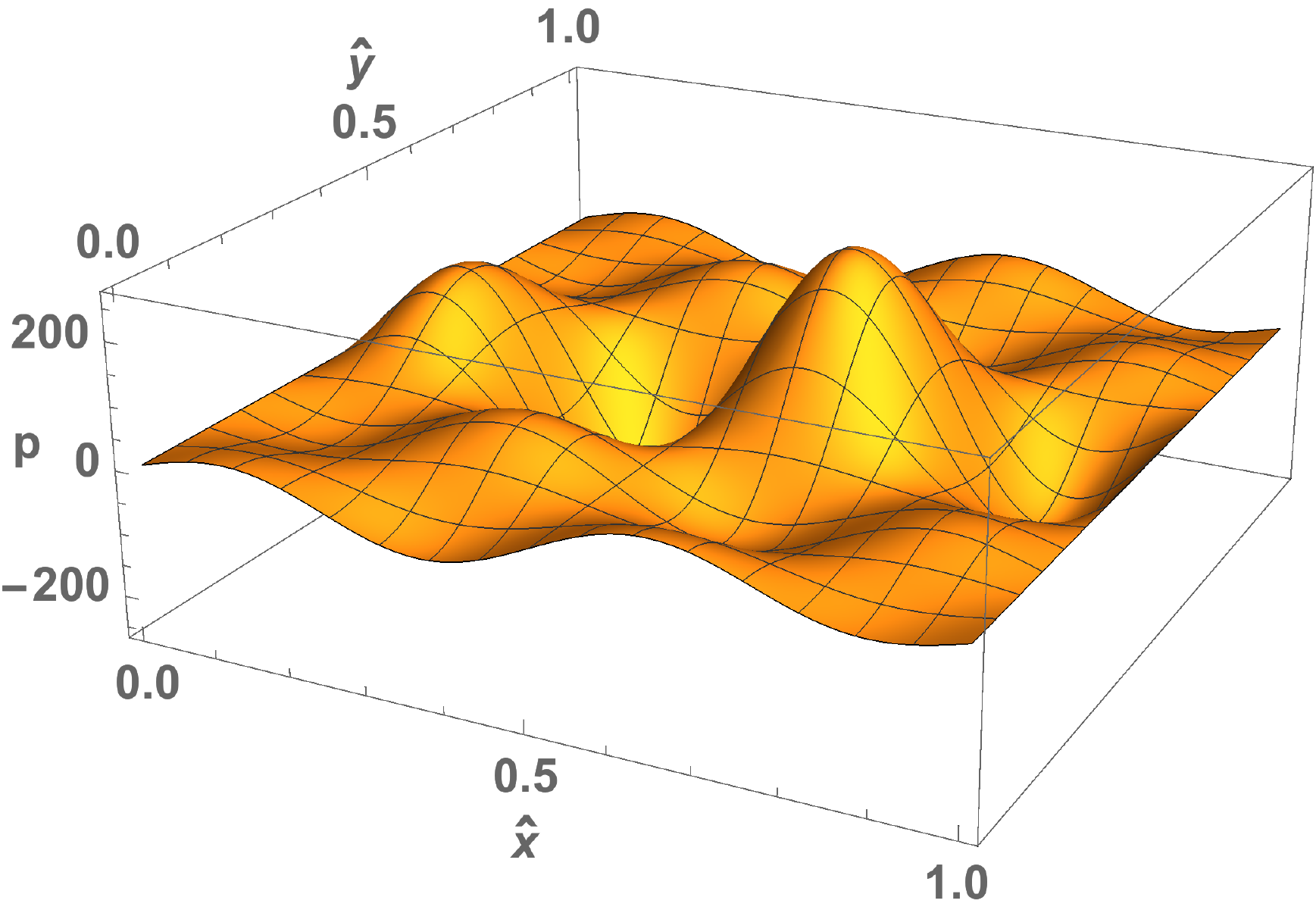}
\caption{Plot of $\hat Q^i$ and $p$ corresponding to corresponding to the metric deformation $\Phi$ 
as in \eqref{fdef} with $\alpha=1$ and $\beta=6.6\times 10^{-4}$
and thermal gradient just in the $\hat x$ direction given by $\hat\zeta = d\hat x$.
The left plot displays the vector heat current density $\hat Q^i$, for twice the period in both spatial directions.
The background colour depicts the norm of the vector, $(\hat Q^i\hat Q^j\delta_{ij})^{1/2}$, and we note that it is nearly uniform with the 
maximum and minimum varying by about 1\%. The right plot shows $p$ for a single period in the spatial directions. The variation in $p$ is uniform enough to lead to a roughly homogenous response.
\label{secondone}}
\end{figure}

For small values of $1-\alpha$ and $\beta$, we are not only in the hydrodynamic limit, we are also in the perturbative limit that is associated with small amplitudes as discussed in \cite{Donos:2015gia,Banks:2015wha}. In this limit,
at leading order in a perturbative expansion in the
amplitude of the metric deformation around flat spacetime, 
the solutions to the Stokes equations are homogenous, {\it i.e.} constant
\cite{Donos:2015gia,Banks:2015wha}. In figure \ref{secondone} we have plotted the solutions to the Stokes equations for $\alpha=1$ and $\beta=6.6\times 10^{-4}$. As expected we find nearly homogeneous flows.
There are various ways of quantifying this: for example the approximate range of components of the current
are $\hat Q^1\in(4565,4595)$ and $\hat Q^2\in (-8.498,8.498)$. The background colour in 
figure \ref{secondone} depicts the norm of the vector field. The maximum value of the norm (red) has components
$(4595, <10^{-6})$ while the minimum (purple) has components\footnote{The second component, in both cases, converges to zero within our numerical accuracy.} $(4565, <10^{-6})$. To compare with the perturbative lattice analysis of \cite{Donos:2015gia,Banks:2015wha} we let the perturbative parameter, $\lambda$, be
equal to the difference between the maximum and the minimum values of $\Phi$ within one period and we find
$\lambda=0.01$. From the above data we see that, roughly, $\hat{Q}^{1}$ scales like $\lambda^{-2}/2$ while $p$ scales like $2\,\lambda^{-1}$.
In any event, there is no thermal backflow for these lattices.

\begin{figure}[h!]
\centering
\includegraphics[width=0.45\textwidth]{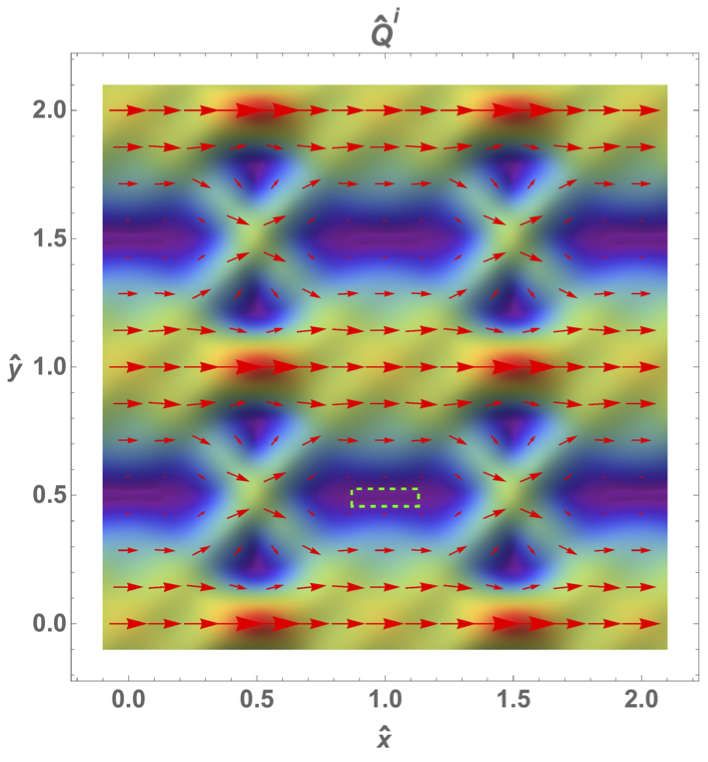}\qquad
\includegraphics[width=0.45\textwidth]{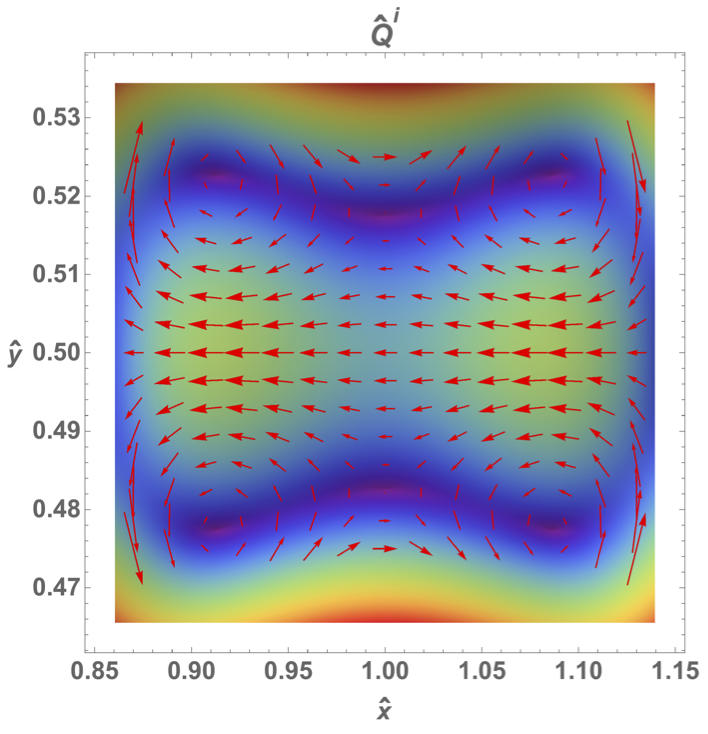}\\
\includegraphics[width=0.45\textwidth]{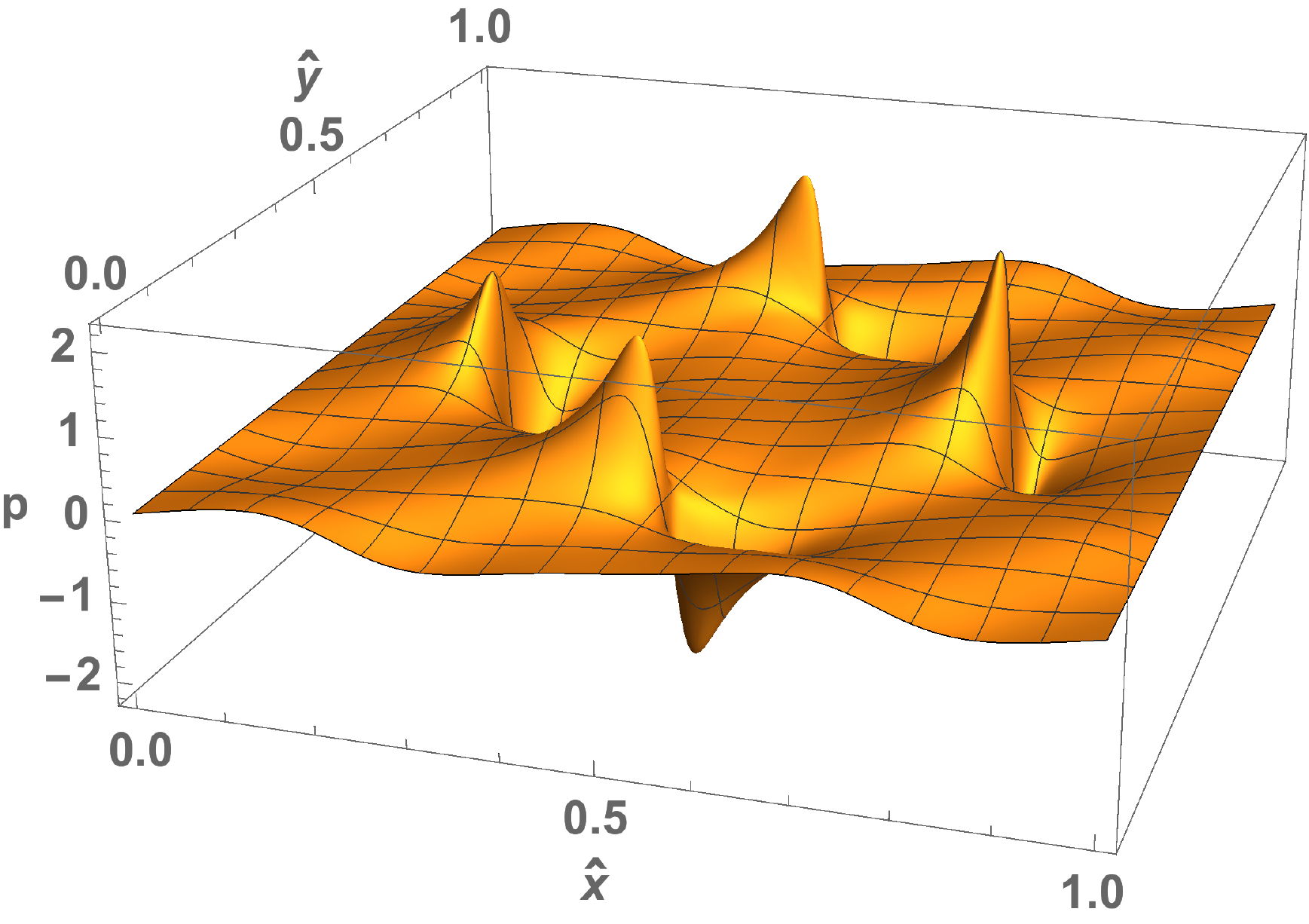}
\caption{Thermal backflow corresponding to the metric deformation in figure \ref{defphi}, 
with $\Phi$ as in \eqref{fdef} with $\alpha=0.98$, $\beta=0.3$, 
and thermal gradient just in
the $\hat x$ direction given by $\hat\zeta = d\hat x$.
The upper plots display the vector heat current density $\hat Q^i$, with the background colour emphasising
the norm of the vector $(\hat Q^i\hat Q^j\delta_{ij})^{1/2}$. The upper left plot shows $\hat Q^i$ for twice the period in both spatial directions. Thermal
backflow occurs in the elongated purple regions: the upper right plot is an enlargement of the green dashed rectangle.
The bottom plot displays $p$ for a single period in the spatial directions.
\label{defphiresults}}
\end{figure}

By increasing the overall amplitude, by varying $\alpha,\beta$, we find that solving the Stokes equations gives rise to sharper peaks in $p$, 
which are associated with larger internal fluid forces. We find that for amplitude fixed by $\alpha=0.98$ and $\beta=0.3$,
that thermal backflow does indeed occur as shown in figure \ref{defphiresults}.
In particular, we 
see that there is a distinct region of thermal backflow with $\hat y\sim 0.5$ and 
$0.8< \hat x< 1.2$

Finally, it is worth revisiting the original assumptions concerning our hydrodynamic expansion with $\epsilon<<1$. 
Recall that throughout this paper we have been assuming the constitutive relation given in \eqref{eq:EMTe}. This will receive corrections
at higher order in $\epsilon$ and will include terms involving the curvature of the background metric.
For the specific example
with $\alpha=0.98$ and $\beta=0.3$ we can estimate that the next order curvature contributions will be of the order
$\epsilon^2$ times $\Phi^{-1}\nabla^2 \ln \Phi$. Since the latter has spikes of the order $10^5$, in order to ensure that
these terms are indeed sub-leading we should impose not just $\epsilon<<1$ but the stricter bound $\epsilon<<10^{-3}$. It would be interesting to determine by how much this can be weakened for other examples exhibiting backflow.

\section{Discussion}
By solving a system of Stokes equations we have shown that thermal backflow driven by an applied 
external DC thermal source
is possible for CFTs in the leading order viscous hydrodynamic limit.  We explicitly demonstrated this for CFTs defined
on static spacetime metrics with a conformally flat spatial metric, with the conformal factor depending periodically on the spatial coordinates. 
We did not have to make any assumption concerning the strength of the viscosity $\eta$ in \eqref{eq:EMTe}; we only demanded that it is non-zero.
The thermal backflow occurs at the level of linear response, and is associated with specific two
point functions of the stress tensor in the CFT. The thermal backflow solutions are steady state solutions
to the linearised equations. If one was interested in going beyond linear response, then
one would have to take into account Joule heating and there would not be such steady state solutions. It would be interesting to understand the time scale for when the linearised approximation breaks down.

We have discussed in section \ref{sec1} how thermal transport properties of CFTs are invariant under Weyl transformations. 
This means, for example, that since backflow occurs if suitable isotropic strains are applied to a CFT, associated with a conformally flat metric $\Phi dx^i dx^i$, then we should also see exactly the same backflow by applying
a periodic local temperature profile parametrised by $\Phi^{-1}$ with a flat spatial metric $dx^i dx^i$. Thus, if one were able to experimentally engineer
such isotropic strains and local temperature profiles for some strongly coupled matter and one found the same thermal response, this
would provide a sharp diagnostic that the matter was described by a conformal field theory in the hydrodynamic limit. Perhaps it is
possible to investigate this with graphene, which is known to be described as a relativistic fluid at the Dirac point.

For general CFTs it is straightforward to generalise our analysis from static to stationary metrics. This corresponds to allowing for deformations of the CFT which have sources for local rotations
in thermal equilibrium as discussed in \cite{Donos:2015bxe}. The linear response to applying a thermal source can then
be examined in the leading order viscous hydrodynamic limit by studying Navier-Stokes equations that contain Coriolis terms which are determined by the non-vanishing vorticity tensor of the background fluid in thermal equilibrium. In the case of DC thermal sources the relevant 
time independent Stokes equations were given in \cite{Donos:2015bxe}. In general it is necessary to focus on the transport currents, which are obtained by subtracting off certain magnetisation currents that depend 
on the applied thermal source \cite{PhysRevB.55.2344,Hartnoll:2007ih,Blake:2015ina,Donos:2015bxe}.

In this paper we have discussed the DC response of general CFTs in the leading order viscous hydrodynamic limit, by solving a system of Stokes equations. For the special class of CFTs that have holographic duals we can also study DC response for deformed CFTs far from the hydrodynamic limit, by analysing suitable black hole solutions. It is a remarkable fact that the total thermoelectric current fluxes, and hence the thermoelectric DC conductivities, can be obtained by solving the same system of Stokes equations for an auxiliary fluid on the horizon of the black holes \cite{Donos:2015gia,Banks:2015wha,Donos:2015bxe}. The connection with hydrodynamic limit was explained in \cite{Banks:2016krz}. Another interesting direction would be to use holography to examine what happens to the backflow as a function of $\epsilon=k/T$.

We can also generalise the analysis in this paper to CFTs that have additional conserved currents. 
From the work on holography \cite{Banks:2016krz} we can conclude that we will need to solve the Stokes equations presented in \cite{Donos:2015gia,Banks:2015wha,Donos:2015bxe}. There is a range of possibilities to examine, including the role of charge puddles and
magnetic fields, and we aim to report on some of this soon.

We have also presented the equations needed to be solved to examine the thermal response
for a general relativistic quantum field theory in appendix \ref{gencase}. For the special case of DC response, for background spacetimes in which the norm of the timelike Killing vector is constant, the relevant equations are, up to constants, the same Stokes equations that need to be solved for the
case of CFTs. In particular, the examples of thermal backflow that we showed in section 3
are applicable to a much more general class of quantum field theories. When the norm of the Killing vector is not constant,
the equations that need to be solved are given in \eqref{tcomp},\eqref{scomp} and it would
be interesting to explore them in more detail.

\section*{Acknowledgements}
We thank Mike Blake for prompting us to include the material in appendix A.
The work of JPG, TG and LM is supported by the European Research Council under the European Union's Seventh Framework Programme (FP7/2007-2013), ERC Grant agreement ADG 339140. The work of JPG is also supported
by STFC grant ST/L00044X/1 and EPSRC grant EP/K034456/1. JPG is also supported as a KIAS Scholar and
as a Visiting Fellow at the Perimeter Institute. The work of EB supported by an Imperial College Schr\"odinger Scholarship.

\appendix

\section{General quantum field theories}\label{gencase}
We now consider a general relativistic quantum field theory, relaxing the constraint of conformal invariance.
The set-up is very similar to that in section 2 and we again use the material in \cite{Baier:2007ix}. We now just impose the Ward identity $D_\mu T^{\mu}{}_{\nu}=0$. 
For the constitutive relation we write
\begin{align}
T_{\mu\nu}= P g_{\mu\nu}+(\varepsilon+P)\,u_{\mu}u_{\nu}+\tau_{\mu\nu}\,,
\end{align}
where 
\begin{align}
\tau_{\mu\nu}=-2\eta\sigma_{\mu\nu} -\zeta_b(g_{\mu\nu}+u_\mu u_\nu)D_\rho u^\rho\,,
\end{align}
Here $\sigma_{\mu\nu}$ is the same as in \eqref{shear} and $\zeta_b$ is the bulk viscosity
and should not be confused with the external thermal source one-form $\zeta=\zeta_\mu dx^\mu=d\phi$.
For CFTs we have $\zeta_b=0$. We also have the local thermodynamic relation
and first law, which take the form
\begin{align}\label{eost}
\varepsilon+P=Ts\,,\qquad dP=s\, dT\,.
\end{align}

To simplify the presentation, we will again just consider static backgrounds with Killing vector $\partial_t$. 
As we will see, background metrics with $\partial_t$ having non-constant norm, {\it i.e.} $g_{tt}\equiv-f^2$ non-constant, will play an interesting role. In considering the perturbation about the background
we note that $P,\varepsilon, S,\eta$ and $\zeta_b$ are all functions of the local temperature. They can depend
on other dimensionful parameters, but these will all be held fixed in the perturbations we are interested in.
Thus, we can write $\varepsilon_0\equiv\varepsilon(T_0)$, $\delta\varepsilon\equiv(\partial_T\varepsilon)_0\delta T$ etc. 
For the perturbed metric and fluid velocity we write
\begin{align}\label{dcpert2}
ds^{2}&=-f^2(x)(1-2\phi)\,dt^{2}+g_{ij}(x)dx^{i}dx^{j}\,,\nn
u_t&=-f(x)(1-\phi),\qquad u_j=\delta u_j\,.
\end{align}
where $\phi$ and $\delta u_i$ are both functions of $(t,x)$ as before. A calculation then gives the stress tensor
\begin{align}
T_{tt}&=\varepsilon_{0}\,f^{2}\left(1-2\phi\right)+\delta\varepsilon\,f^{2}\,,\nn
T_{ti}&=-f\,(\varepsilon_0+P_{0})\,\delta u_{i}\,,\nn
T_{ij}&=(P_{0}+\delta P)\,g_{ij}-2\eta_0\,f^{-1}\,\nabla_{(i}\left(f\delta u_{j)}\right)
+\left(\frac{2\eta_0}{(d-1)}-\zeta_{b0}\right) g_{ij}\,f^{-1}\,\nabla_{k}(f\,\delta u^{k})\,.
\end{align}
The heat current, defined in \eqref{qexp} is given by
\begin{align}
Q^i=\sqrt{g}f^2(\varepsilon_{0}+P_{0})\delta u^{i}=\sqrt{g}f^2T_0s_0\delta u^{i}\,.
\end{align}

We next note that in order to ensure that the Ward identity is 
satisfied for the unperturbed background we must have 
\begin{align}\label{widnp}
f^{-1}\partial_i f(\varepsilon_0+P_0)+\nabla_i P_0=0\,.
\end{align}
Using the equation of state and the first law in \eqref{eost} for the background we can then
integrate \eqref{widnp} to find
\begin{align}\label{teeeq}
T_0=\bar T_0 f^{-1}\,,
\end{align}
where $\bar T_0$ is a constant. In particular, we see that in general $T_0$ depends on the spatial coordinates.

Returning now to the perturbed stress tensor, for the time component of the Ward identity we obtain
\begin{align}\label{tcomp}
f\partial_{t}\delta\varepsilon+\nabla_{i}(f^{2}(\varepsilon_{0}+P_{0})\delta u^{i})=0\,.
\end{align}
For the spatial component, and using \eqref{widnp}, we find 
\begin{align}\label{scomp}
&f^{-1}(\varepsilon_{0}+P_{0})\,\partial_{t}\delta u_{i}+f^{-1}\partial_i f(\delta\varepsilon+\delta P)-(\varepsilon_0+P_0)\zeta_i+\partial_{i}\delta P\nn
&\qquad-2f^{-1}\,\nabla^{j}\left(\eta_0\,\nabla_{(j}\left(f\delta u_{i)}\right)\right)+f^{-1}\,\nabla_{i}\left( \left( \frac{2\eta_0}{d-1}-\zeta_{b0}\right) \nabla_{k}(f\delta u^{k})\right)=0\,.
\end{align}
Notice that the time component of the four-vector $\zeta_t$ again does not appear. The perturbations 
$\delta \varepsilon$ and $\delta P$ can both be expressed in terms of $\delta T$ since we are holding all other
dimensionful parameters fixed. In fact, using \eqref{eost} we have 
$\delta P=s_0\delta T$ and $\delta\varepsilon=T_0(\partial_Ts)_0\delta T$. 
Thus these equations should again be solved for $\delta T$ and $\delta u_i$.

When $f=1$, from \eqref{teeeq} we have that $T_0$ is a constant. As a consequence 
$P_0,\varepsilon_0,s_0,\eta_0$ and
$\zeta_{b0}$ are then also constants. 
In this case the Ward identities simplify
to the following linearised Navier-Stokes equations
\begin{align}
&T_0^{-1}\partial_{t}\delta T+c_s^2\nabla_{i}\delta u^{i}=0\,,\nn
&T_0s_0\,\partial_{t}\delta u_{i}+s_0\partial_{i}\delta T
-2\eta_0 \,\nabla^{j}\nabla_{(j}\delta u_{i)}+
\left( \frac{2\eta_0}{d-1}-\zeta_{b0}\right) \nabla_{i} \nabla_{k}\delta u^{k}=T_0s_0\zeta_i\,.
\end{align}
In the first equation we have introduced the speed of sound squared, 
$c_s^2=(\partial_\epsilon P)_0 =s_0/(T_0(\partial_Ts)_0)$. For a CFT we have $c_s^2=1/(d-1)$.
Moreover, to study DC response we can set the time derivatives to zero and we obtain the Stokes equations
for an incompressible fluid
\begin{align}
\nabla_{i}\delta u^{i}=0\,,\qquad
\partial_{i}\delta T
-2\frac{\eta_0}{s_0} \,\nabla^{j}\nabla_{(j}\delta u_{i)}
=T_0\zeta_i\,.
\end{align}

\section{Numerical integration}\label{appnum}

We want to solve the system of equations \eqref{stokesgeneral2} for the variables
$v_i$, $p$ for a specified constant $\hat \zeta_i$ on
a torus with unit periods and metric $g_{ij}$.
In order to numerically solve this boundary value problem for a two dimensional horizon, we will discretise our domain on $N_{x}\times N_{y}$ points. Given the periodicity of the problem and the fact that we expect to find smooth solutions, we use Fourier pseudo-spectral methods to approximate the derivatives of our functions on our computational grid. The problem then reduces to a $(3\,N_{x}\,N_{y})\times(3\,N_{x}\,N_{y})$ inhomogeneous linear system which we can write in matrix form as
\begin{align}
\mathbb{M}\cdot \mathbf{v}=\mathbf{s}\,.
\end{align}
The $3\,N_{x}\,N_{y}$ dimensional vector $\mathbf{v}$ is used to store the values of the functions $p$, $v_{x}$ and $v_{y}$ on the grid. In more detail
\begin{align}
\mathbf{v}_{i}=\begin{cases}
p_{i \bmod{N_{x}},\left[\tfrac{i}{Nx} \right]},& 1\leq i \leq  N_{x} N_{y}\\
(v_{x})_{(i-N_{x}N_{y}) \bmod N_{x},\left[\tfrac{i-N_{x}N_{y}}{Nx} \right]},& N_{x} N_{y} <i \leq 2 N_{x} N_{y}\\
(v_{y})_{(i-2N_{x}N_{y}) \bmod N_{x},\left[\tfrac{i-2N_{x}N_{y}}{Nx} \right]},& 2 N_{x} N_{y}<i \leq 3 N_{x} N_{y}
\end{cases}
\end{align}
where $\left[\frac{a}{b}\right]$ denotes the integer part of the division between $a$ and $b$. The vector $\mathbf{s}$ is reserved for the inhomogeneous part of system \eqref{stokesgeneral2} and it does depend on the direction of the temperature gradient. For example, when the temperature gradient is just along the $x$ direction, and unit valued, we have
\begin{align}
(\mathbf{s}^{x})_{i}=\begin{cases}
1,& N_{x} N_{y}<i \leq 2 N_{x} N_{y}\\
0, &\mathrm{otherwise}
\end{cases}
\end{align}
 It is easy to see that we only have to do a single inversion of the matrix $\mathbb{M}$ and the solution for sources in different directions can simply be found by a matrix multiplication of $\mathbb{M}^{-1}$ with the corresponding source vector $\mathbf{s}^{x}$ or $\mathbf{s}^{y}$.

We have implemented the method outlined above in \verb!C++! taking advantage of the language's templates to write code which can be used with various data types. However, we found that \verb!double! precision was enough to obtain accurate solutions for our purposes. We did find though that we had to use quite large resolutions of the order of $N_{x}=N_{y}\sim 181$. This need is becoming obvious from our plots since there is small scale features we have to resolve. One example is the sharp peaks in the plots of $p$. The linear solver we used was the version of \verb!PARDISO! included with Intel's \verb!MKL! \verb!BLAS! suite. The specific solver can take advantage of \verb!OpenMP! at several stages of the solution of the linear system which proved useful when we ran our code on multicore systems.


\providecommand{\href}[2]{#2}\begingroup\raggedright\endgroup

\end{document}